\documentclass[sigconf, natbib=true]{acmart}

\usepackage{xcolor}
\usepackage[linesnumbered,ruled,vlined]{algorithm2e}

\SetCommentSty{mycommfont}
\SetKwInput{KwInput}{Input}                % Set the Input
\SetKwInput{KwOutput}{Output}              % set the Output

\newcommand\blfootnote[1]{%
  \begingroup
  \renewcommand\thefootnote{}\footnote{#1}%
  \addtocounter{footnote}{-1}%
  \endgroup
}

\newcommand{\expnumber}[2]{{#1}\mathrm{e}{#2}}

\usepackage{enumitem}
\newlist{questions}{enumerate}{2}
\setlist[questions,1]{label=RQ\arabic*.,ref=RQ\arabic*}
\setlist[questions,2]{label=(\alph*),ref=\thequestionsi(\alph*)}

\usepackage{setspace}

\usepackage{mathtools}

\usepackage{multirow}
\usepackage{array}
\usepackage{subcaption}
\AtBeginDocument{%
  \providecommand\BibTeX{{%
    \normalfont B\kern-0.5em{\scshape i\kern-0.25em b}\kern-0.8em\TeX}}}

\setcopyright{acmcopyright}
\copyrightyear{2024}
\acmYear{2024}
\acmDOI{XXXXXXX.XXXXXXX}

\acmConference[SIGIR '24]{The 47th International ACM SIGIR Conference on Research and Development in Information Retrieval}{July 14--18, 2024}{Washington, D.C.}
\acmPrice{15.00}
\acmISBN{978-1-4503-XXXX-X/18/06}

\newcommand{\ie}{\textit{i.e.}, } %Literally.
\newcommand{\eg}{\textit{e.g.}, } %For example.
\newcommand{\aka}{\textit{a.k.a. }}
\newcommand{\etal}{\textit{et al.} } %And others, with correct punctuation.
\begin{document}

%%
%% The "title" command has an optional parameter,
%% allowing the author to define a "short title" to be used in page headers.
\title[Lightweight Embeddings for Graph Collaborative Filtering]{Lightweight Embeddings for Graph Collaborative Filtering}

%%
%% The "author" command and its associated commands are used to define
%% the authors and their affiliations.
%% Of note is the shared affiliation of the first two authors, and the
%% "authornote" and "authornotemark" commands
%% used to denote shared contribution to the research.
% \author{
%     Xurong Liang{\small$^\dag$}\hspace*{10pt}Tong Chen{\small$^\dag$}\hspace*{10pt}Lizhen Cui{\small$^\perp$}\hspace*{10pt}Yang Wang{\small$^\S$}\hspace*{10pt}Meng Wang{\small$^\S$}\hspace*{10pt}Hongzhi Yin{\small$^{\dag *}$}\\
%     \fontsize{10}{10}\selectfont\itshape $~^\dag$The University of Queensland, Australia, {\fontsize{9}{9}\selectfont\ttfamily\upshape \{xurong.liang,tong.chen,h.yin1\}@uq.edu.au}\\
%     \fontsize{10}{10}\selectfont\itshape $~^\perp$Shandong University, China, {\fontsize{9}{9}\selectfont\ttfamily\upshape clz@sdu.edu.cn}\\
%     \fontsize{10}{10}\selectfont\itshape $^\S$Hefei University of Technology, China, {\fontsize{9}{9}\selectfont\ttfamily\upshape yangwang@hfut.edu.cn, eric.mengwang@gmail.com	}
% }

\author{Xurong Liang}
\email{xurong.liang@uq.edu.au}
\orcid{0000-0002-3458-3887}
\affiliation{%
  \institution{The University of Queensland}
  \city{Brisbane}
  % \state{Queensland}
  \country{Australia}
  % \postcode{4072}
}

\author{Tong Chen}
\email{tong.chen@uq.edu.au}
% \orcid{0000-0002-3458-3887}
\affiliation{%
  \institution{The University of Queensland}
  \city{Brisbane}
  \country{Australia}
}

\author{Lizhen Cui}
\email{clz@sdu.edu.cn}
\affiliation{%
  \institution{Shandong University}
  \city{Jinan}
  \country{China}
}

\author{Yang Wang}
\email{yangwang@hfut.edu.cn}
\affiliation{%
  \institution{Hefei University of Technology}
  \city{Hefei}
  \country{China}
}

\author{Meng Wang}
\email{eric.mengwang@gmail.com}
\affiliation{%
  \institution{Hefei University of Technology}
  \city{Hefei}
  \country{China}
}

\author{Hongzhi Yin*}
\email{h.yin1@uq.edu.au}
\orcid{0000-0003-1395-261X}
\affiliation{%
  \institution{The University of Queensland}
  \city{Brisbane}
  \country{Australia}
}

%%
%% By default, the full list of authors will be used in the page
%% headers. Often, this list is too long, and will overlap
%% other information printed in the page headers. This command allows
%% the author to define a more concise list
%% of authors' names for this purpose.
\renewcommand{\shortauthors}{Xurong Liang \etal}

%%
%% The abstract is a short summary of the work to be presented in the
%% article.
\begin{abstract}
Graph neural networks (GNNs) are currently one of the most performant and versatile collaborative filtering methods. Meanwhile, like in traditional collaborative filtering, owing to the use of an embedding table to represent each user/item entity as a distinct vector, GNN-based recommenders have inherited its long-standing defect of parameter inefficiency. As a common practice for scalable embeddings, parameter sharing enables the use of fewer embedding vectors (which we term meta-embeddings), where each entity is represented by a unique combination of meta-embeddings instead. When assigning meta-embeddings, most existing methods are a heuristically designed, predefined mapping from each user/item entity's ID to the corresponding meta-embedding indexes (\eg double hashing), thus simplifying the optimization problem into learning only the meta-embeddings. However, in the context of GNN-based collaborative filtering, such a fixed mapping omits the semantic correlations between entities that are evident in the user-item interaction graph, leading to suboptimal recommendation performance. To this end, we propose \textbf{L}ightweight \textbf{E}mbeddings for \textbf{G}raph \textbf{C}ollaborative \textbf{F}iltering (LEGCF), a parameter-efficient embedding framework dedicated to GNN-based recommenders.
LEGCF innovatively introduces an assignment matrix as an additional learnable component on top of meta-embeddings. To jointly optimize these two heavily entangled components, aside from learning the meta-embeddings by minimizing the recommendation loss, LEGCF further performs efficient assignment update by enforcing a novel semantic similarity constraint and finding its closed-form solution based on matrix pseudo-inverse. The meta-embeddings and assignment matrix are alternately updated, where the latter is sparsified on the fly to ensure negligible storage overhead. Extensive experiments on three benchmark datasets have verified LEGCF's smallest trade-off between size and performance, with consistent accuracy gain over state-of-the-art baselines. The codebase of LEGCF is available in \url{https://github.com/xurong-liang/LEGCF}. 

\end{abstract}

%%
%% The code below is generated by the tool at http://dl.acm.org/ccs.cfm.
%% Please copy and paste the code instead of the example below.
%%
% \begin{CCSXML}
% <ccs2012>
%  <concept>
%   <concept_id>10010520.10010553.10010562</concept_id>
%   <concept_desc>Computer systems organization~Embedded systems</concept_desc>
%   <concept_significance>500</concept_significance>
%  </concept>
% \end{CCSXML}

% \ccsdesc[500]{Computer systems organization~Embedded systems}
% \ccsdesc[300]{Computer systems organization~Redundancy}
% \ccsdesc{Computer systems organization~Robotics}
% \ccsdesc[100]{Networks~Network reliability}

%%
%% Keywords. The author(s) should pick words that accurately describe
%% the work being presented. Separate the keywords with commas.

%\keywords{Lightweight Recommender Systems, Compositional Embeddings, Graph Collaborative Filtering}

%%
%% This command processes the author and affiliation and title
%% information and builds the first part of the formatted document.
\maketitle

\blfootnote{*Hongzhi Yin is the corresponding author.}

\section{Introduction} \label{sec:intro}
In e-commerce and social media applications, the success of recommender systems (RSs) in predicting preferred items for users \cite{wu_graph_2023, zhang_deep_2020, zheng_automl_2023, zhang2021double, wang2017location, zhao2023survey} has attracted immense revenue. 
Classic RSs deploy factorization-based methods \cite{rendle_factorization_2010,koren_matrix_2009} or multilayer perceptrons (MLPs) \cite{he_neural_2017} for predicting user-item affinity via collaborative filtering. Recent advances in graph neural networks (GNNs) \cite{wu_graph_2023,wang2019neural,he_lightgcn_2020} further facilitate graph-based collaborative filtering \cite{wang2019neural,he_lightgcn_2020}, which propagates collaborative signals through connected nodes to model user-item interactions. 
To date, GNN-based recommenders are arguably one of the most popular variants of collaborative filtering methods in both academia \cite{wang2019neural,he_lightgcn_2020} and industry \cite{ying2018graph, gurukar2022multibisage}. 

While GNN-based recommenders bring in promising performance and versatility, they still struggle to bypass a long-standing challenge carried over from conventional RSs -- that is, the scalability of embedding representations. In the very basic form of ID-based recommendation where the only entities\footnote{In this paper, we term both users and items as entities for convenience.} to represent are users and items, 
each entity corresponds to an individual embedding vector that can be looked up from the embedding table based on a unique feature value (\eg ID). With the sheer volume of users and items in e-commerce sites, the embedding table inevitably introduces a huge number of parameters that render a recommender hard to scale. The embedding parameters are at the scope of tens of millions in common benchmark datasets \cite{chen_learning_2021, xia_-device_2022, liang2023learning} and even thousands of billions in industry-level applications \cite{mudigere2022software}. Thus, to keep enjoying the performance benefits of GNN-based recommenders at scale, improving the parameter efficiency of embeddings is the key.

To alleviate the excessive parameterization in the embedding layer of RSs, one mainstream research direction focuses on parameter sharing, in which all entity embeddings are drawn from a parameter pool with only a fraction of the footprint of the full embedding table. 
As a typical way of parameter sharing, \textit{compositional embeddings} are facilitated via a set of meta-embedding vectors, where some hash functions are handcrafted to map each entity to a unique set of meta-embeddings based on their feature value (\ie user/item IDs in common RSs) \cite{weinberger_feature_2009, zhang_model_2020, yan2021binary, das2007google, desai2021semantically}. 
Then, an entity embedding is composed with those meta-embeddings, \eg by performing sum pooling \cite{shi_compositional_2020}. As each entity is assigned a distinct set of meta-embeddings, their final compositional embeddings naturally avoid collisions and can be easily used for subsequent recommendation tasks.   

Nevertheless, in these methods, it is noticed that the mapping from entities to their corresponding meta-embeddings is solely based on the hash functions given \cite{shi_compositional_2020,li2021lightweight}, which in turn makes all entities' meta-embedding assignments a predefined and fixed decision. Although this narrows the optimization target of this lightweight embedding paradigm down to just the meta-embeddings, such an assignment mapping unfortunately retains minimal semantic association between entities since it is solely determined by the user/item IDs that are mutually independent. That is to say, even two dissimilar entities, \eg a user and her disliked item, might have two similar meta-embedding compositions where most meta-embeddings are the same, resulting in a misleadingly high affinity between their composed embeddings. Similarly, two highly correlated entities can hardly preserve such correlations in their embeddings if they diverge heavily in meta-embedding assignments. 
In the meantime, most compositional embedding schemes allocate a unanimous weight to all meta-embeddings \cite{liang2023learning,lian_lightrec_2020} with no mechanisms to differentiate their contributions in an entity's compositional embedding, indulging the homogeneity of entity embeddings. 
As a result, risks are incubated in terms of insufficient embedding expressiveness and suboptimal recommendation performance, and they are likely to deteriorate in GNN-based collaborative filtering where user/item links are explicitly accounted for message passing.

To this end, a more flexible compositional embedding scheme that not only accounts for the semantic connections among entities, but also provides a more nuanced, weighted composition of assigned meta-embeddings for each entity, is highly anticipated. Ideally, this involves learning the right meta-embedding assignments as a parallel goal to the optimization of meta-embeddings. However, this also means selecting a subset of meta-embeddings for every entity, translating into a typical combinatorial optimization problem that is known to be NP-hard. At the same time, for each entity embedding, any update on either its meta-embedding assignments or the meta-embedding themselves will incur changes in its expressiveness. Given the strong entanglement between the two, simultaneously learning them is prone to instability during optimization and heavily constrained recommendation efficacy. 

In this paper, to achieve the desiderata of parameter-efficient embeddings with GNN-based recommenders, we propose our work \textbf{L}ightweight \textbf{E}mbeddings for \textbf{G}raph \textbf{C}ollaborative \textbf{F}iltering (\textbf{LEGCF}).
Specifically, to remedy the defects of predefined hash function methods, we propose a more adaptive assignment scheme by introducing a learnable, real-valued assignment matrix, of which each row encodes which (selection) and how (weighting) every meta-embedding is involved in the composition of one entity's embedding. To learn the assignment matrix while preserving the semantic relationships among entities, we make use of GNNs' ability to propagate messages between associated nodes. Specifically, we construct an expanded message passing graph that additionally considers the weighted connections between entities and meta-embeddings, and frame a representation similarity constraint as an objective function that can be elegantly and quickly solved in a closed form via matrix pseudo-inverse. This not only transforms the combinatorial optimization task into inferring the relevance of different meta-embeddings to each entity, but also decouples it from the gradient-based update of meta-embedding weights. Furthermore, in LEGCF, dynamic pruning is enabled for the assignment matrix, such that it can be efficiently stored as a sparse matrix with a negligible memory footprint.

We summarize our main contributions as follows:
\begin{itemize}
    \item We point out the deficiencies of existing embedding parameter sharing methods when they are applied in GNN-based recommenders. Instead of using predefined meta-embedding assignments, we highlight the necessity of a flexible meta-embedding assignment mechanism that can preserve entity-wise semantic correlations and differentiate the nuanced contributions of each entity's assigned meta-embeddings.
    \item To achieve the desiderata, we propose a novel lightweight embedding framework dedicated to GNN-based recommenders, namely LEGCF. On top of learning all meta-embeddings, LEGCF adaptively learns weighted meta-embedding assignment for each entity. A novel, closed-form solution is derived from the similarity among graph-propagated embeddings, so as to facilitate an alternate optimization scheme.
    \item We conduct extensive experiments on three benchmark datasets. The results have demonstrated LEGCF's advantageous accuracy amid its top-tier parameter efficiency among state-of-the-art lightweight baselines, and verified its smallest performance compromise compared with GNN-based recommenders using a full (40$\times$ larger) embedding table. 
\end{itemize}

\section{Related Work} \label{sec:related_work}
In this section, we present the background of real-valued embedding compression in recommender systems.

\subsection{Parameter Sharing}
A large proportion of work achieves embedding compression by allowing multiple entities to share embeddings from a meta-embedding pool
and hashing is the most notable trick for this purpose. 
These methods often utilize multiple hash functions to gather several meta-embeddings and compose them together to form entity embeddings (\ie compositional embedding) to avoid hash collision. 

\textbf{Compositional Embedding Methods.}
Zhang \etal \cite{zhang_model_2020} and Shi \etal \cite{shi2020beyond} adopt multiple hashing to generate composed embeddings for parts of the entities.
Q-R trick \cite{shi_compositional_2020, li2021lightweight} deploys 
the quotient and remainder codebooks to generate complementary composed embeddings. Yan \etal \cite{yan2021binary} propose to break the binary code of ID value into multiple small hash values, each of which yields a meta-embedding for the entity. SCMA \cite{desai2021semantically} relies on locality sensitive hashing (LSH) to allocate segments of a single memory chunk to entities. Embeddings generated by tensor train (TT) decomposition methods \cite{xia_-device_2022, yin2021tt, yin_nimble_2022, wang_next_2020} can be treated as an alternative form of compositional embedding, as
every entity embedding is a multiplication product of a series of tensors.
Quantization \cite{zhang2022anisotropic, lian_lightrec_2020, liu2021online, liu2020online, lian2020product, jiang2021xlightfm} methods
find the best fit codes from multiple codebooks and compose them to form entity embeddings, which is also compositional.

Apart from compositional embedding methods, some work facilitates the goal of parameter sharing in other ways. DHE replaces the hash embedding table with a deep neural network (DNN) \cite{kang_learning_2021}. LCE \cite{hang2022lightweight} only stores the embeddings of users/items and the embeddings of their counterparts are inferred at training time. Alternatively, clustering can be a strategy for entity meta-embedding assignment \cite{das2007google}. CEL \cite{chen2023clustered} takes a different approach to split one giant cluster into multiple small clusters based on entity similarity. Entities in each cluster have a unified embedding.

Despite all methods mentioned above generating a memory-efficient embedding layer, no work leverages the user-item interaction graph for meta-embedding allocation. Hence, there is no direct mapping between entity interaction and representation proximity in the latent space, potentially leading to inadequately learned entity embeddings and hindering the downstream task.

\subsection{AutoML Embedding Optimization}
Another hot spot of embedding optimization studies lies in the application of automatic machine learning (AutoML) \cite{zheng_automl_2023, yin2024device}. 
One main direction of research is the automated dimension search \cite{zhao_autoemb_2020, liu_automated_2020, qu2023continuous, cheng2020differentiable, ginart2021mixed, joglekar_neural_2020}, in which a set of hand-crafted entity embedding candidate dimension sizes is supplied beforehand. The algorithms can then devise a differentiable dimension search \cite{zhao_autoemb_2020, cheng2020differentiable}, or a reinforcement learning agent \cite{liu_automated_2020, qu2023continuous, joglekar_neural_2020, qu2024budgeted} to determine the embedding dimension size for each entity. Despite these methods being effective in finding the balance between memory budget and the recommendation performance, 
the low computational efficiency of NAS \cite{pham2018efficient} and the huge search space for reinforcement learning often raise concern. Pruning is another popular direction \cite{liu_learnable_2021, kong_autosrh_2022, deng2021deeplight, li2023dual, yao2022razor, yan2021learning, chen_learning_2021, lyu_optembed_2022, qu2022single, liang2023learning, zheng2024personalized}. 
Work that purely relies on pruning to reach the memory target \cite{liu_learnable_2021, deng2021deeplight, yao2022razor, yan2021learning, qu2022single} may face the challenge of null-value embedding vector for entities, in which all elements lied in the embedding vector are zero, severely impacting recommendation performance. 
Later work \cite{kong_autosrh_2022, li2023dual, chen_learning_2021, lyu_optembed_2022} attempts to combat the negative impact of pruning technique by combining other embedding optimization strategies, such as AutoML dimension search or embedding hashing. However, these hybrid methods do not overcome the inherited drawback of low computational efficiency,  making it intractable to large-scale deployment.

% \section{Preliminaries} \label{sec:preliminaries}
% \input{sections/preliminaries.tex}

\section{Method} \label{sec:method}
\begin{figure*}[t!] 
    \centering
    \begin{subfigure}[b]{.35\textwidth}
            \centering
            \includegraphics[width=.95\textwidth, height=3.1cm]{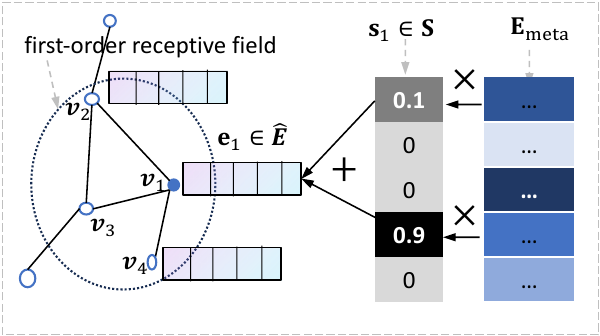}
            \vspace*{-2mm}
            \caption[]%
            {{\small Compositional entity embeddings}}    
    \end{subfigure}%
    \begin{subfigure}[b]{.2\textwidth}
            \centering
            \includegraphics[width=.95\textwidth, height=3.1cm]{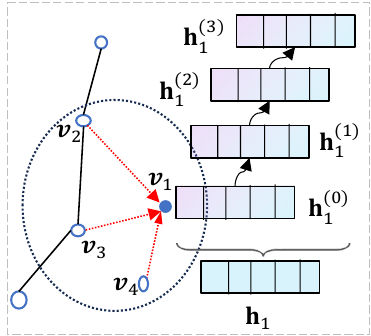}
            \vspace*{-2mm}
            \caption[]%
            {{\small Graph propagation}}    
    \end{subfigure}%
    \begin{subfigure}[b]{.2\textwidth}
            \centering
            \includegraphics[width=.97\textwidth, height=3.1cm]{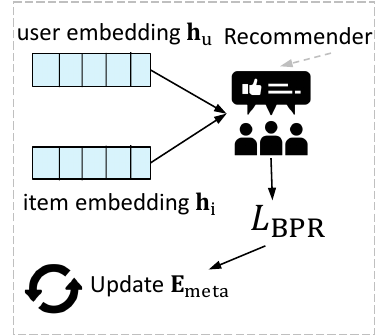}
            \vspace*{-2mm}
            \caption[]%
            {{\small Learning $\textbf{E}_{meta}$}}    
    \end{subfigure}%
    \begin{subfigure}[b]{.2\textwidth}
            \centering
            \includegraphics[width=.95\textwidth, height=3.1cm]{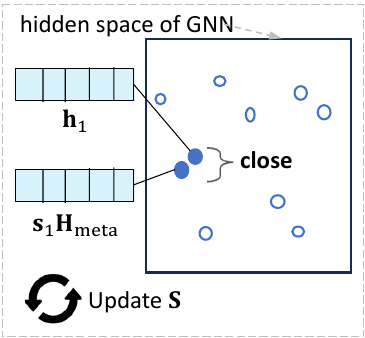}
            \vspace*{-2mm}
            \caption[]
            {{\small Learning \textbf{S}}}    
    \end{subfigure}%

    \vspace*{-4mm}
    \caption{Key components of LEGCF. 
    Corresponding details can be found in Section \ref{sec:comp_emb} for (a), Section \ref{sec:meta_emb_learning} for (b) and (c), and Section \ref{sec:weight_update} for (d). Note that we use $\textbf{e}$ and $\textbf{s}$ to denote one row of the codebook $\textbf{E}_{meta}$ and assignment matrix $\textbf{S}$, respectively.}
    \label{fig:overview}

    \vspace*{-4mm}
\end{figure*}

We now present LEGCF, our proposed lightweight embedding method for GNN-based recommenders. In Figure \ref{fig:overview}, we provide visual aids regarding LEGCF's key components, whose details are introduced in this section. 

\subsection{Replacing Full Embedding Table with Compositional Embeddings}\label{sec:comp_emb}
We denote the sets of users and items as $\mathcal{U}$ and $\mathcal{I}$, respectively. 
As we focus on ID-based recommendation in this paper, the total number of entities is $N = |\mathcal{U}| + |\mathcal{I}|$. The user-item interactions are encoded in a binarized matrix $\textbf{R} \in \{0, 1\}^{|\mathcal{U}| \times |\mathcal{I}|}$, indicating whether an interaction is observed between a user-item pair. To minimize the memory footprint of the embedding layer, the full embedding table $\textbf{E} \in \mathbb{R}^{N \times d}$ with embedding dimensionality $d$ is replaced by a memory-efficient meta-embedding codebook $\textbf{E}_{meta} \in \mathbb{R}^{c \times d}$, where and $c\ll N$ is the number of meta-embeddings, \aka bucket size. To compose an embedding for each entity, a common approach \cite{li2021lightweight,shi_compositional_2020,liang2023learning} is to hash an entity's index $p$ into $t$ non-overlapping meta-embedding indexes $\{q_1, q_2, ..., q_t\}$, based on which the corresponding meta-embeddings are drawn from the codebook $\textbf{E}_{meta}$. Each entity's embedding is then formed by merging those $t$ meta-embeddings, usually via pooling or element-wise product. 

In our work, we abandon the use of fixed hash functions and instead introduce a learnable assignment matrix $\textbf{S} \in \mathbb{R}^{N \times c}_{\geq 0}$. Essentially, each row of the assignment matrix $\textbf{S}$ corresponds to a user/item entity, where the non-zero elements indicate which meta-embeddings are selected out of the $c$ choices, as well as their weights in the composition of the entity embedding. The entries in $\textbf{S}$ are updated during training, providing high flexibility in embedding generation. 
With the meta-embedding codebook $\textbf{E}_{meta}$ and learnable assignment matrix $\textbf{S}$, all entities' compositional embeddings $\widehat{\textbf{E}} \in \mathbb{R}^{N \times d}$ can be computed via the following matrix multiplication:
\begin{equation} \label{eq:get_full_embs}
    \widehat{\textbf{E}} = \textbf{S} \textbf{E}_{meta}.
\end{equation}
Notably, considering the size of a dense assignment matrix, we constrain $\textbf{S}$ to be highly sparse, where corresponding details are deferred to Section \ref{sec:weight_update}.

In GNN-based recommenders, a user-item interaction graph is built to propagate collaborative signals among connected entities for embedding learning \cite{wu_graph_2023}, and a common practice \cite{he_lightgcn_2020,wang2019neural} is to represent those connections via an adjacency matrix $\textbf{A} \in \mathbb{R}^{N \times N}$:
\begin{equation} \label{eq:original_graph}
    \textbf{A} = \begin{bmatrix} \textbf{0}& \textbf{R}\\\textbf{R}^\top&\textbf{0} \end{bmatrix},
\end{equation}
where $\textbf{R}$ is the user-item interaction matrix.
For each user $u\in \mathcal{U}$ and item $i\in \mathcal{I}$, we first retrieve their embeddings $\textbf{e}_u$ and $\textbf{e}_i$ from $\widehat{\textbf{E}}$, and treat them as the inputs (\ie layer-0 embeddings) to GNN. 
In short, each GNN layer involves the propagation of embeddings over the graph $\textbf{A}$, where the final graph-propagated embeddings $\textbf{h}_u$, $\textbf{h}_i\in \mathbb{R}^d$ are computed by applying a pooling operation to the embeddings of each $u$/$i$ obtained from each GNN layer. In our paper, we use LightGCN \cite{he_lightgcn_2020} as the base recommender and follow its configuration to use mean pooling over all layers' propagated embeddings as the final representation of each entity. 
As our main innovation lies in the new compositional embedding paradigm for GNN-based recommenders, we omit other details of LightGCN for simplicity. For each ($u$, $i$) pair, their recommendation affinity prediction can be calculated as the dot product of their graph-propagated embeddings:
\begin{equation} \label{eq:rating_computation}
    \widehat{y}_{ui} = \textbf{h}_u^\top \textbf{h}_i,
\end{equation}
which facilitates the Bayesian personalized ranking loss (BPR) \cite{rendle_bpr_2009} for optimizing each entity's associated meta-embeddings:
\begin{equation} \label{eq:bpr_loss}
    \mathcal{L}_{\textit{BPR}} = \sum_{(u, i^+, i^-) \in \mathcal{B}} - \ln{\sigma(\widehat{y}_{ui^+}  - \widehat{y}_{ui^-})} + \lambda ||\Theta||^2,
\end{equation}
where $\mathcal{B}$ is either the whole training set or a training batch, $(u, i^+, i^-)$ is a triplet that contains sampled user $u$'s observed interacted item $i^+$ and unvisited item $i^-$, $||\Theta||^2$ is the $L_2$ regularization over trainable parameters and $\lambda$ controls its weight in the loss.

\textbf{Batch Processing on Full Embedding Table Inference.}
Note that while we define the forward pass for computing all entities' embeddings as a unified matrix operation in Eq.~\ref{eq:get_full_embs}, in cases where a large user and item number $N$ challenges memory usage during training, our compositional embeddings also supports batch processing like in standard GNN-based recommenders. To do so, in the $b$-th batch, we can draw $n\ll N$ rows from $\textbf{S}$, forming a $n \times c$ matrix $\textbf{S}_{b}$ for $n$ entities. As such, for the $b$-th batch, Eq.~\ref{eq:get_full_embs} can be rewritten as $\widehat{\textbf{E}}_{b} = \textbf{S}_{b} \textbf{E}_{meta}$ with $\widehat{\textbf{E}}_{b}\in \mathbb{R}^{n\times c}$, which is compatible with all the subsequent computations.

\subsection{Graph-Propagated Meta-Embeddings} \label{sec:meta_emb_learning}
The effectiveness of GNNs in refining entity embeddings via message passing has been widely acknowledged \cite{wu_graph_2023, gao2023survey}. In LEGCF, the compositional embedding table $\widehat{\textbf{E}}$ is generated from Eq. \ref{eq:get_full_embs} with the meta-embedding codebook $\textbf{E}_{meta}$. As each meta-embedding is essentially linked to a set of nodes within the user-item interaction graph $\textbf{A}$, we represent such connections by incorporating the assignment matrix $\textbf{S}$ into $\textbf{A}$: 
\begin{equation} \label{eq:A_expanded}
    \textbf{A}^{\prime} = \begin{bmatrix} \textbf{A}& \textbf{S}\\\textbf{S}^\top&\textbf{0} \end{bmatrix},
\end{equation}
where $\textbf{A}^{\prime} \in \mathbb{R}^{(N+c) \times (N+c)}$ is termed an \textit{expanded interaction graph}. Intuitively, $\textbf{A}^{\prime}$ appends meta-embeddings as additional virtual nodes to the user-item interaction graph, thus taking advantages of the GNN for $\textbf{E}_{meta}$ to propagate.
The graph propagation of meta-embeddings resembles the same message passing process described in Section \ref{sec:comp_emb}, where the input embeddings (\ie when layer $l=0$) to GNN $\textbf{H}^{(0)} \in \mathbb{R}^{(N + c) \times d}$ is generated by stacking the compositional embeddings $\widehat{\textbf{E}}$ and the meta-embedding codebook $\textbf{E}_{meta}$:
\begin{equation} \label{eq:H_0}
    \textbf{H}^{(0)} = \begin{bmatrix}
        \widehat{\textbf{E}}\\ \textbf{E}_{meta}
    \end{bmatrix}.
\end{equation}
The propagation operation at layer $l+1$ is thus defined as:
\begin{equation} \label{eq:H_new_layer}
        \textbf{H}^{(l+1)} = (\textbf{D}^{-\frac{1}{2}} \textbf{A}^{\prime} \textbf{D}^{-\frac{1}{2}}) \textbf{H}^{(l)}, 
\end{equation}
where $\textbf{A}^\prime$ is the adjacency matrix from the expanded interaction graph, $\textbf{D}^\prime \in \mathbb{R}^{(N + c) \times (N + c)}$ is the diagonal degree matrix of $\textbf{A}^\prime$. $\textbf{D}^{-\frac{1}{2}} \textbf{A}^\prime \textbf{D}^{-\frac{1}{2}}$ creates a symmetrical, degree-normalized adjacency matrix. The final GNN embeddings can be gathered by taking the mean of embeddings from all layers:
\begin{equation} \label{eq:H_final}
    \textbf{H} = \frac{1}{L + 1} \sum_{l = 0}^{L} \textbf{H}^{(l)},
\end{equation} 
from which the graph-propagated representations of the full entity set $\textbf{H}_{full}\in \mathbb{R}^{N\times d}$ and meta-embeddings $\textbf{H}_{meta}\in \mathbb{R}^{c\times d}$ can then be respectively retrieved by splitting $\textbf{H}$ into the following form:
\begin{equation} \label{eq:get_h_full_h_meta}
        \textbf{H}_{full} \gets \textbf{H}[:N, \textit{ }:], \,\,\,
        \textbf{H}_{meta} \gets \textbf{H}[N:, \textit{ }:].
\end{equation}
Specifically, the graph-propagated meta-embeddings $\textbf{H}_{meta}$ are used for assignment matrix update as described in Section \ref{sec:weight_update}. In general, as $\textbf{H}_{meta}$ now conceals collaborative information from semantically similar user/item neighbors, it facilitates learning a more sensible meta-embedding assignment for each user/item. 
To conduct the recommendation task, we retrieve the entities' graph-propagated embeddings from $\textbf{H}_{full}$, and then calculate the similarity score between each user-item pair as defined in Eq. \ref{eq:rating_computation}. With a fixed assignment $\textbf{S}$ in each iteration, the predicted scores of a training batch are passed to Eq. \ref{eq:bpr_loss} to calculate the BPR loss $\mathcal{L_{\textit{BPR}}}$, thus enabling back-propagation to update the values in the meta-embedding codebook $\textbf{E}_{meta}$. 

\subsection{Learning Assignment Weights} \label{sec:weight_update}
While the meta-embedding codebook $\textbf{E}_{meta}$ is being learned, the weights in $\textbf{S}$ should be updated over time as well in pursuit of a more reasonable mapping between entities and their meta-embedding assignments. As discussed earlier, a sensible meta-embedding assignment for each entity should be able to account for its own semantic relations to other entities. That is to say, two closely correlated entities are expected to have their embeddings composed by a similar set of meta-embeddings and vice versa -- an intuitive analogy to the core idea behind collaborative filtering. 

In this regard, a straightforward solution is to directly update the weights within $\textbf{S}$ via gradient descent in the same back-propagation loop with $\textbf{E}_{meta}$. However, as the recommendation loss is directly affected by their product $\widehat{\textbf{E}}$, when simultaneously updating the weights in both $\textbf{S}$ and $\textbf{E}_{meta}$, the co-adaptation between their updates can introduce difficulties in finding a stable solution \cite{hinton2012improving,larsson2016fractalnet}. Also, for each entity, instead of assigning weights to all $c$ meta-embeddings and thus making $\textbf{S}$ a dense matrix, a more desirable objective is to make it highly sparse, \ie for every row of $\textbf{S}$, only a few entries are non-zero. On the one hand, it makes less sense to associate each entity with all meta-embeddings in a compositional paradigm. On the other hand, if the $N\times c$ assignment matrix is dense, even a moderate choice of $c$ (\eg $c=100$) can make the storage cost of $\textbf{S}$ comparable to the $N\times d$ full embedding table (\eg $d=128$ as a popular choice), which defeats the purpose of compressing it at the first place. As a possible path to the discrete, weighted selection of meta-embeddings, reinforcement learning \cite{sutton2018reinforcement} unfortunately requires a specifically crafted action sampler and reward function, and the optimization of meta-embedding combinations from a vast action space is hardly tractable.  

To resolve the challenges, recall that $\textbf{S}$ defines the mapping from each entity to its corresponding meta-embeddings in the input layer of the GNN, where each $\textbf{S}[p,q] \neq 0 $ can be viewed as a similarity score between the $p$-th entity and the $q$-th meta-embedding. Optimally, if such similarity holds in the input layer, it should still be preserved between the entity embeddings and meta-embeddings after $L$ layers of propagation. 
Therefore, we describe this property as the following semantic similarity constraint:
\begin{equation} \label{eq:assignment_mat_in_hidden_space}
    \textbf{H}_{full} \approx \textbf{S} \textbf{H}_{meta},
\end{equation}
where $\textbf{H}_{full}$ and $\textbf{H}_{meta}$ are the graph-propagated embeddings as in Section \ref{sec:meta_emb_learning}. Assuming $\textbf{H}_{full}$ and $\textbf{H}_{meta}$ are in optimal form, the assignment matrix $\textbf{S}$ which reflects the relationship between $\textbf{H}_{full}$ and $\textbf{H}_{meta}$ can be solved as:
\begin{equation} \label{eq:get_S_by_inv}
    \textbf{S} = \textbf{H}_{full} \textbf{H}^{-1}_{meta},
\end{equation}
with $\textbf{H}^{-1}_{meta} \in\mathbb{R}^{d\times c}$ being the inverse of $\textbf{H}_{meta} \in\mathbb{R}^{c\times d}$. However, as it is impractical to assume $c=d$, $\textbf{E}_{meta}$ is commonly a non-square matrix, neither does $\textbf{H}_{meta}$. Consequently, the direct inverse of $\textbf{H}_{meta}$ may not exist. As a wraparound, we approximate $\textbf{H}^{-1}_{meta}$ with the pseudo-inverse of $\textbf{H}_{meta}$. 
We resort to the Moore-Penrose inverse \cite{ben2003generalized} and solve the pseudo-inverse of $\textbf{H}_{meta}$ through singular value decomposition (SVD) \cite{stewart1993early} due to its computational simplicity and accuracy \cite{ben2003generalized}. To be specific, we first decompose $\textbf{H}_{meta}$ into the following form:
\begin{equation}
    \textbf{H}_{meta} = \textbf{U} \textbf{$\Sigma$} \textbf{V}^{*},
\end{equation}
where $\textbf{U} \in \mathbb{R}^{c \times c}$ is a unitary matrix, $\Sigma \in \mathbb{R}^{c \times c}$ is the square matrix that contains singular values along the diagonal, $\textbf{V}^{*} \in \mathbb{R}^{c \times d}$ is the conjugate transpose of unitary matrix $\textbf{V} \in \mathbb{R}^{d \times c}$. The Moore-Penrose pseudo-inverse of $\textbf{H}_{meta}$, denoted as $\textbf{H}^{ \dagger}_{meta} \in \mathbb{R}^{d \times c}$, is computed as:
\begin{equation} 
    \textbf{H}^{ \dagger}_{meta} = \textbf{V} \Sigma^{-1} \textbf{U}^{*},
\end{equation}
where $\Sigma^{-1} \in \mathbb{R}^{c \times c}$ is the inverse matrix of $\Sigma$, which can be easily obtained by replacing the singular values along the diagonal of $\Sigma$ with their reciprocal, $\textbf{U}^{*} \in \mathbb{R}^{c \times c}$ is the conjugate transpose of $\textbf{U}$. Once the pseudo-inverse of $\textbf{H}_{meta}$ is calculated, we can plug it back into Eq. \ref{eq:get_S_by_inv} to find the closed-form solution for $\textbf{S}$:
\begin{equation} \label{eq:update_assignment_mat}
    \textbf{S} = \textbf{H}_{full} \textbf{H}^{ \dagger}_{meta} = \textbf{H}_{full} \textbf{V} \Sigma^{-1} \textbf{U}^{*}.
\end{equation}
In this way, utilizing the constraint defined in Eq. \ref{eq:assignment_mat_in_hidden_space}, we can derive the updated assignment matrix $\textbf{S}$ via matrix pseudo-inverse. Such a gradient-free learning strategy prevents gradient-tracking on $\textbf{S}$ along with $\textbf{E}_{meta}$, and is very efficient to compute. Moreover, it can effectively preserve the entities' semantic associations in their meta-embedding assignments, as two similar rows in $\textbf{H}_{full}$ will lead to similar results after the multiplication in Eq. \ref{eq:update_assignment_mat}. 

\subsection{Sparsifying, Batching, and Initializing \textbf{S}} \label{sec:matrix_init}
As a continuation of Section \ref{sec:weight_update}, we follow up with more design details about learning the assignment matrix $\textbf{S}$.

\textbf{Sparsification on Assignment Matrix.} 
Since Eq. \ref{eq:update_assignment_mat} yields a dense, real-valued matrix, to keep $\textbf{S}$ sparse, a sparsification process is required to make its storage footprint manageable. Fortunately, our gradient-free assignment weight update strategy allows us to modify $\textbf{S}$ in place, where we conduct sparsification to retain the largest $t$ non-zero weights in each row of $\textbf{S}$ so that the number of meta-embeddings used by each entity is kept at $t$ at all time:
\begin{equation} \label{eq:assignment_mat_topk}
    \textbf{S}[p,q] = \begin{cases} 
      \textbf{S}[p,q] & \text{if}\,\, q \in \text{idx}_{\text{top-}t}(\textbf{S}[p,:]) \\
      0 & \text{otherwise}
   \end{cases},
\end{equation}
where $\text{idx}_{\text{top-}t}(\cdot)$ returns the indexes of the top-$t$ entries of a given vector (\ie the $p$-th row of \textbf{S} in our case). 
Essentially, only for meta-embeddings considered critical for an entity embedding, we retain their relatively larger weights in $\textbf{S}$. In the worst case that every meta-embedding in $\textbf{E}_{meta}$ is selected in the sparsified $\textbf{S}$, LEGCF exerts a space complexity of $O(tN + cd)$ for its lightweight embeddings. Considering $t\ll c$ ($t\leq 5$ in our settings), $O(tN + cd)\ll O(Nd)$ compared to the space complexity of a full embedding table, implying a significant improvement in parameter efficiency.

\textbf{Batch Processing on Assignment Weight Update.}
In a similar vein to Eq. \ref{eq:get_full_embs}, Eq. \ref{eq:assignment_mat_in_hidden_space} also supports batch computing, which can be enabled if lower memory consumption during training is desired. This is also facilitated by slicing the assignment matrix $\textbf{S}$ into batches, each of which has $n$ rows ($n \ll N$). The operation in Eq. \ref{eq:update_assignment_mat} to update $\textbf{S}$ is thus performed batch-wise:
\begin{equation}
        \textbf{S}_b = \textbf{H}_b \textbf{H}^{\dagger}_{meta},
\end{equation}
where $b$ indexes the $b$-th batch. 
This substantially lowers the in-memory space complexity from $O(Nc + Nd + cd)$ to $O(nc + nd + cd)$, where the computed $n$ dense rows of $\textbf{S}$ are immediately passed into Eq. \ref{eq:assignment_mat_topk} to obtain its sparse version. 

\textbf{Assignment Matrix Initialization.} 
Given that similar entities are expected to own similar meta-embeddings, a good initialization of assignment matrix $\textbf{S}$ can potentially contribute to higher recommendation accuracy and a speedup in training.  
In $\textbf{S}$, for each entity $p$, we term its highest-weighted meta-embedding (among the $t$ selected) the \textit{anchor meta-embedding} indexed by $q^*_{p}$, and it forms the base image of the entity's embedding $\textbf{e}_p$ since it takes the largest proportion in its composition.
To let correlated entities have similar base embedding images during initialization, we propose to assign two entities $p_1$, $p_2$ the same anchor meta-embedding weight if they demonstrate high affinity with each other, \ie $\textbf{S}[p_1, q^*_{p_1}]=\textbf{S}[p_1, q^*_{p_2}]$.
Because the entities lying within each community in fact form tightly connected subgraphs in the original interaction graph, assigning them the same anchor meta-embedding helps pass the same proximity onto the latent space. In LEGCF, we take advantage of a well-established multilevel graph partitioning algorithm, namely METIS \cite{karypis1997metis} for partitioning the user-item interaction graph. Although other advanced graph clustering methods can also be considered, METIS well serves the one-off initialization purpose due to its fast and accurate computation (no learning involved), balanced and non-overlapping partitions (all anchor meta-embeddings are fairly utilized), as well as deterministic results (ease of replication). 
We set the desired partition number to $c$, where entities in the same subgraph share one specific anchor meta-embedding in $\textbf{E}_{meta}$. 
To reflect this, on initialization of the assignment matrix $\textbf{S}$, we perform the following for every entity $p$: 
\begin{itemize}[leftmargin=*]
\item[I.] Given entity $p$'s subgraph index $c_p\in \{1,2,...,c\}$, we set $\textbf{S}[p,q_p^*]= w^*$, where $q_p^*=c_p$ and $w^*$ is a universal hyperparameter.
\item[II.] We uniformly sample $t-1$ indexes from $\{1,2,...,c\}\setminus c_p$ with replacement, denoted as a set $\mathcal{Q}$. For every $q\in \mathcal{Q}$, we set the corresponding assignment weight $\textbf{S}[p,q]=\frac{(1 - w^*)}{t-1}$. 
\item[III.] For all remaining entries at $q'\notin \mathcal{Q}\cup c_p$, we set $\textbf{S}[p,q^{\prime}]=0$.
\end{itemize}
In short, each entity receives one initial anchor meta-embedding based on the subgraph assigned, while the remaining $t-1$ meta-embedding assignments are randomly initialized in favor of diversity. The impact of $w^*$ will be examined later as a hyperparameter.

\subsection{The Overall Algorithm}
\vspace*{-6mm}

\begin{algorithm}
\begin{spacing}{.8}
\small
\renewcommand{\arraystretch}{.80}

    \caption{The algorithm of LEGCF.}\label{alg:algo}
        
        Randomly initialize $\textbf{E}_{meta} \in \mathbb{R}^{c \times d}$
        
        Initialize $\textbf{S} \in \mathbb{R}^{N \times c}$ following Section \ref{sec:matrix_init}
        
        Compute $\textbf{A}^{\prime} \in \mathbb{R}^{(N + c) \times (N + c)}$ with Eq. \ref{eq:A_expanded}
        
        \tcc{Pretraining}
        \While{not converged}{Optimize $\mathcal{L}_{BPR}$ w.r.t. Eq. \ref{eq:A_expanded}---\ref{eq:get_h_full_h_meta}}

        \tcc{Assignment Update Enabled}
        epoch\_index $\gets 0$
        
        \While{not converged}{
            Optimize $\mathcal{L}_{BPR}$ w.r.t. Eq. \ref{eq:A_expanded}---\ref{eq:get_h_full_h_meta}

            \tcc{$m$ is a hyperparameter for update frequency}
            \If{epoch\_index $\textit{mod } m == 0$}{
                Perform assignment update with Eq. \ref{eq:update_assignment_mat}

                Regenerate $\textbf{A}^{\prime}$ following Eq. \ref{eq:A_expanded}
            }

            epoch\_index $\gets$ epoch\_index $+ 1$
        }
\end{spacing}
\end{algorithm}
\vspace*{-6mm}

As depicted by the pseudocode in Algorithm \ref{alg:algo}, our proposed framework LEGCF consists of two stages. The first stage is the warm-up pretraining stage, wherein we randomly initialize the meta-embedding codebook $\textbf{E}_{meta}$, generate the assignment matrix $\textbf{S}$ using METIS graph partitioning, and precompute the extended adjacency matrix $\textbf{A}'$ (lines 1-3). We freeze \textbf{S} and only perform the downstream recommendation task to provide a stable environment for meta-embedding codebook learning (lines 4-5). 
Once the learning of $\textbf{E}_{meta}$ converges, we introduce the assignment matrix update to learn entities' assignment weights in $\textbf{S}$ and further refine the meta-embedding codebook $\textbf{E}_{meta}$ by alternating between the two procedures (lines 6-12). Note that as the meta-embedding assignments are computed with the graph-propagated representations, we only update $\textbf{S}$ every $m$ epochs to ensure the graph-propagated representations are fairly stable by the time of updating $\textbf{S}$. Furthermore, when the assignment matrix is updated, the expanded graph adjacency matrix $\textbf{A}^{\prime}$ is also regenerated (line 11) to reflect updates in connections between entities and meta-embeddings.

\section{Experiments} \label{sec:experiments}
In this section, we conduct experiments to validate the effectiveness of our work. We break down our experiments into the following research questions (RQs):
\begin{itemize}
    \item \textbf{RQ1:} How is the recommendation accuracy of our work compared to other baselines under a tight memory budget?
    \item \textbf{RQ2:} Does METIS initialization and learnable assignment update improve recommendation performance?
    \item \textbf{RQ3:} How sensitive is our work to hyperparameters?
    \item \textbf{RQ4:} How does using different GNN recommenders affect our work's accuracy?
\end{itemize}

\subsection{Experimental Settings}

\subsubsection{Datasets}
We use three publicly available benchmark datasets: \textbf{Gowalla}, \textbf{Yelp2020} and \textbf{Amazon-book}, which can be found in the official code repositories of \cite{he_lightgcn_2020} and \cite{sun2021hgcf}. The detailed statistics of datasets are summarized in Table \ref{tab:dataset_stat}. 

\subsubsection{Base Recommender and Baselines}
As described in Section \ref{sec:comp_emb}, the GNN backbone of LEGCF is implemented based on LightGCN \cite{he_lightgcn_2020}. 
For evaluation fairness, we also adopt LightGCN as the base recommender in the following baselines: 
\begin{itemize}
    \item \textbf{Variable-Size Methods} Methods in this category allows the generation of embedding layers to satisfy various memory targets. This is accomplished by applying pruning (\textbf{PEP} \cite{liu_learnable_2021}) or changing the hash embedding structure (\textbf{QR} \cite{shi_compositional_2020}). Note that our work is also capable of generating embedding layers of various memory usage. This is done by modifying the number of meta-embedding buckets $c$ and the number of composition embeddings assigned to each entity $t$. 

    \item \textbf{One-Size Methods} Methods in this category do not formalize the final memory target as an optimization objective. Thus, it is impractical to ensure the final embedding structure meets the memory budget. Baselines that belong to this category include AutoML-based dimension search algorithms \textbf{AutoEmb} \cite{zhao_autoemb_2020}, \textbf{ESAPN} \cite{liu_automated_2020}, \textbf{OptEmbed} \cite{lyu_optembed_2022}, \textbf{CIESS} \cite{qu2023continuous}; \textbf{DHE} \cite{kang_learning_2021} which replaces embedding hashing with DNN network; and \textbf{NimbleTT} \cite{yin_nimble_2022, yin2021tt} which approximates the conventional embedding table using tensor-train decomposition. We omit the result of \textbf{CEL} \cite{chen2023clustered} in this section due to its poor performance in our empirical testing.
\end{itemize}
Apart from the above baselines, to compare the performance compromise of various lightweight embedding methods, we also conduct experiments on the non-compressed versions of LightGCN with relatively large dimension sizes (\ie 64 and 128), which we term the Unified Dimensionality (\textbf{UD}) setting.

\begin{table}[t!]
\renewcommand{\arraystretch}{.85}
    \caption{Statistics of datasets used in our work.}
    \label{tab:dataset_stat}
    \centering
     \vspace*{-4mm}
    \begin{tabular}{c  c c  c c}
        \toprule
        Dataset & \#User & \#Item & \#Interactions & Density\\
        \midrule
        Gowalla & 29,858 & 40,981 & 1,027,370 & 0.084\%\\
        Yelp2020  & 71,135  & 45,063 & 1,782,999 & 0.056\%\\
        Amazon-book & 52,643 & 91,599 & 2,984,108 & 0.062\%\\
        \bottomrule
    \end{tabular}
    % \vspace*{-5mm}
\end{table}

\subsubsection{Evaluation Metrics}
we select \textbf{NDCG@N} and \textbf{Recall@N} with $N$ set to $\{10, 20\}$ as the evaluation metrics \cite{wang2013theoretical}.

\subsubsection{Implementation Details} \label{sec:implementation_detail}
We follow the train/test/validation protocol specified in \cite{wang2019neural} for dataset splitting. To generate interaction samples for training, for each user-positive item interaction, we randomly draw 5 negative items to form the training set. 
We set the dimension size of embeddings $d$ to $128$. We initialize the meta-embeddings and recommender weights using Xavier Initialization \cite{glorot2010understanding} and then deploy ADAM optimizer \cite{kingma2014adam} for value learning. The learning rate is chosen from $\{\expnumber{1}{-2}, \expnumber{1}{-3}, \expnumber{1}{-4}\}$. The $L_2$ penalty factor $\lambda$ is selected from $\{0, \expnumber{5}{-4}, \expnumber{1}{-3}, \expnumber{5}{-3}, \expnumber{1}{-2}\}$. We use a 3-layer GNN for Gowalla and a 4-layer GNN for Yelp2020 and Amazon-book. The number of compositional embeddings per entity $t$ is set to $2$ to reduce the memory usage of the assignment matrix. On initialization of $\textbf{S}$, the weight of anchor embedding $w^*$ for all entities is searched from the range $\{0.5, 0.6, 0.7, 0.8, 0.9\}$ for each dataset. When the pretrain stage is done, we update the assignment matrix $\textbf{S}$ once every epoch. We choose the default number of meta-embedding buckets in the codebook $c$ to be 500 and reveal the overall performance evaluated under this setting in RQ1. For variable-size baselines, we first calculate the total number of parameters used by our work under the 500-bucket setting, then we use that as the memory budget to select the correct sparsity target/number of hashing buckets for PEP and QR respectively, so that the parameter sizes of their embedding layers do not exceed the nominated memory budget. For baselines incapable of controlling the final parameter size, we define appropriate search spaces to avoid extreme memory footprint on the embedding layer and meanwhile, provide flexibility for these methods to select the most optimal setting. For each experiment setting, we select 3 random seeds to obtain results from multiple runs. We record the average results in this section.

\subsection{Overall Performance (RQ1)}
\begin{table*}[t!]
   % \vspace*{-4mm}
    \caption{Performance comparison between LEGCF and baselines. ``\#Param''  indicates the total parameter size of the \underline{embedding layer} of each method. ``UD - dim 128'' and ``UD - dim 64'' are the full embedding table setting with unified dimensions $d=128$ and $d=64$, respectively. In each column, we use bold font to mark the best result achieved by a lightweight embedding method.}
    \label{tab:overall_performance}
    \vspace*{-4mm}
\centering
\setlength\tabcolsep{1.3pt}
\begin{tabular}{c|ccccc|ccccc|ccccc}
\toprule
             & \multicolumn{5}{c|}{Gowalla}                & \multicolumn{5}{c|}{Yelp2020}               & \multicolumn{5}{c}{Amazon-book}             \\ \hline
Method       & \#Param & N@10   & R@10   & N@20   & R@20   & \#Param & N@10   & R@10   & N@20   & R@20   & \#Param & N@10   & R@10   & N@20   & R@20   \\ \hline
UD - dim 128 & 9.07m   & 0.0901 & 0.1101 & 0.1059 & 0.1576 & 14.87m  & 0.0284 & 0.0426 & 0.0382 & 0.0721 & 18.46m  & 0.0172 & 0.0215 & 0.0230 & 0.0367 \\
UD - dim 64  & 4.53m   & 0.0884 & 0.1088 & 0.1041 & 0.1557 & 7.44m   & 0.0274 & 0.0409 & 0.0366 & 0.0687 & 9.23m   & 0.0165 & 0.0204 & 0.0222 & 0.0353 \\ \hline
ESAPN        & 0.78m   & 0.0268 & 0.0273 & 0.0305 & 0.0405 & 1.12m   & 0.0066 & 0.0083 & 0.0087 & 0.0141 & 1.44m   & 0.0045 & 0.0037 & 0.0051 & 0.0057 \\
AutoEmb      & 0.92m   & 0.0273 & 0.0278 & 0.0319 & 0.0429 & 1.51m   & 0.0071 & 0.0089 & 0.0093 & 0.0148 & 1.44m   & 0.0048 & 0.0039 & 0.0059 & 0.0070 \\
OptEmbed     & 1.42m   & 0.0375 & 0.0370 & 0.0405 & 0.0498 & 5.96m   & 0.0085 & 0.0090 & 0.0109 & 0.0159 & 2.11m   & 0.0038 & 0.0032 & 0.0050 & 0.0062 \\
DHE          & 0.85m   & 0.0049 & 0.0051 & 0.0068 & 0.0105 & 1.39m   & 0.0020 & 0.0025 & 0.0028 & 0.0046 & 1.73m   & 0.0012 & 0.0010 & 0.0015 & 0.0019 \\
NimbleTT     & 0.20m   & 0.0259 & 0.0296 & 0.0300 & 0.0431 & 0.28m   & 0.0055 & 0.0073 & 0.0071 & 0.0122 & 0.31m   & 0.0028 & 0.0031 & 0.0037 & 0.0055 \\
CIESS        & 0.29m   & 0.0643 & 0.0742 & 0.0745 & 0.1080 & 0.47m   & 0.0175 & 0.0253 & 0.0234 & 0.0437 & 0.57m   & 0.0021 & 0.0030 & 0.0029 & 0.0054 \\
PEP          & 0.21m   & 0.0638 & 0.0568 & 0.0698 & 0.0807 & 0.30m   & 0.0170 & 0.0191 & 0.0218 & 0.0326 & 0.29m   & 0.0030 & 0.0024 & 0.0037 & 0.0043 \\
QR           & 0.21m   & 0.0354 & 0.0416 & 0.0412 & 0.0606 & 0.30m   & 0.0067 & 0.0092 & 0.0088 & 0.0157 & 0.29m   & 0.0040 & 0.0046 & 0.0053 & 0.0082 \\
LEGCF        & 0.21m   & \textbf{0.0846} & \textbf{0.0979} & \textbf{0.0988} & \textbf{0.1444} & 0.30m   & \textbf{0.0214} & \textbf{0.0310} & \textbf{0.0291} & \textbf{0.0548} &  0.35m       &  \textbf{0.0134}      &   \textbf{0.0156}     &   \textbf{0.0172}     &  \textbf{0.0259}       \\ \bottomrule
\end{tabular}
\vspace*{-2mm}
\end{table*}

The overall performance of LEGCF with meta-embedding bucket size $c = 500$ and baseline methods are shown in Table \ref{tab:overall_performance}. Regarding recommendation performance, on three datasets, LEGCF achieves the most outstanding result across all evaluation metrics compared to all other embedding optimization baselines. This is verified as the performance result of LEGCF exceeds the second best baseline (\ie CIESS for Gowalla and Yelp2020 and AutoEmb for Amazon-book) by a large margin. The performance boost of LEGCF is especially significant in Amazon-book dataset, revealing LEGCF's suitability in big-scale data application. Comparing the parameter sizes achieved by various embedding optimization methods, LEGCF belongs to the tier that yields the lowest parameter sizes for all three datasets. Yet among its variable-size competitors PEP and QR, LEGCF attains the best performance. As for one-size methods, NimbleTT and CIESS attain similar parameter sizes to variable-size methods but their performance is not comparable to ours. Embedding layers generated by OptEmbed, ESAPN and AutoEmb techniques have much higher parameter sizes than ours. DHE yields the worst performance on all datasets, indicating small dimension size hash codes do not improve embedding uniqueness. Our experiment implies that most one-size methods are not designed to work under a tight memory budget.

On the other hand, comparing the result of LEGCF to unified dimensionality (UD) settings, it is observed that our work sacrifices a tidy drop in performance but in return, the memory usage of the embedding table is significantly reduced. Using the result evaluated on the Gowalla dataset as an example, despite applying LEGCF causes performance degradation of $6.5\%$, the parameter size of the embedding table generated by LEGCF is 22$
\times$ smaller than the full embedding table in UD setting with dimension size = 64 and 44$\times$ smaller than that with dimension size = 128. LEGCF demonstrates an optimal tradeoff between memory usage and recommendation performance, such that the resultant embedding layer can easily fit into resource-constrained devices, yet the recommendation performance is barely compromised.

\subsection{Model Component Analysis (RQ2)}
\begin{table}[t!]
\renewcommand{\arraystretch}{.88}

% \vspace*{-4mm}
\caption{Performance comparison between the default setting and settings with particular component modified. The default setting is the one with assignment update and METIS assignment initialization enabled. We use bold font to indicate the best result.}
\centering
\vspace*{-4mm}
\setlength\tabcolsep{1.3pt}
\label{tab:ablation}
\begin{tabular}{cc|cc|cc|cc}
%\hline
\toprule
                                                           &                                                      & \multicolumn{2}{c|}{Gowalla} & \multicolumn{2}{c|}{Yelp2020} &  \multicolumn{2}{c}{Amazon-book} \\ \hline
\begin{tabular}[c]{@{}c@{}}Assignment\\Update\end{tabular} & \begin{tabular}[c]{@{}c@{}}Init\\Method\end{tabular} & N@20          & R@20         & N@20          & R@20    & N@20 & R@20     \\ \hline
on                                                         & METIS                                                & \textbf{0.0988}        & \textbf{0.1444}       & \textbf{0.0291}        & \textbf{0.0548}  & \textbf{0.0172} & \textbf{0.0259}     \\
off                                                        & METIS                                                & 0.0901        & 0.1254       & 0.0272        & 0.0497  &  0.0147 & 0.0227    \\
on                                                         & random                                               & 0.0788        & 0.1113       & 0.0183        & 0.0327   &   0.0109 & 0.0163  \\ 
%\hline
\bottomrule
\end{tabular}
\end{table}
The innovative components in LEGCF include METIS assignment matrix initialization and the assignment update process. To verify their effectiveness in improving performance, we conduct ablation studies on these two main components. The performance comparison is depicted in Table \ref{tab:ablation}. We further carry out case studies to showcase the learning of entity embeddings and assignment weights in different training stages.

\textbf{Community-based Assignment Initialization.}
We modify the initialization strategy of assignment matrix $\textbf{S}$ to randomly assign anchor embeddings to entities. Our experiment reveals that the settings without METIS initialization perform much worse than those with it. 
This means having a meaningful assignment matrix on initialization is the key step to train an accurate meta-embedding codebook for high-quality entity embeddings.   

\textbf{Assignment Update.}
We eliminate the assignment update process by evaluating the recommender immediately after the pretrain stage converges. 
The drop in performance without assignment update shows the necessity of updating $\textbf{S}$ as it provides the opportunity for entities to alter the compositional weights to generate a more meaningful and customized embedding. 

\begin{figure}
% \vspace*{-6mm}
        \centering
            \centering
            \includegraphics[width=.45\textwidth]{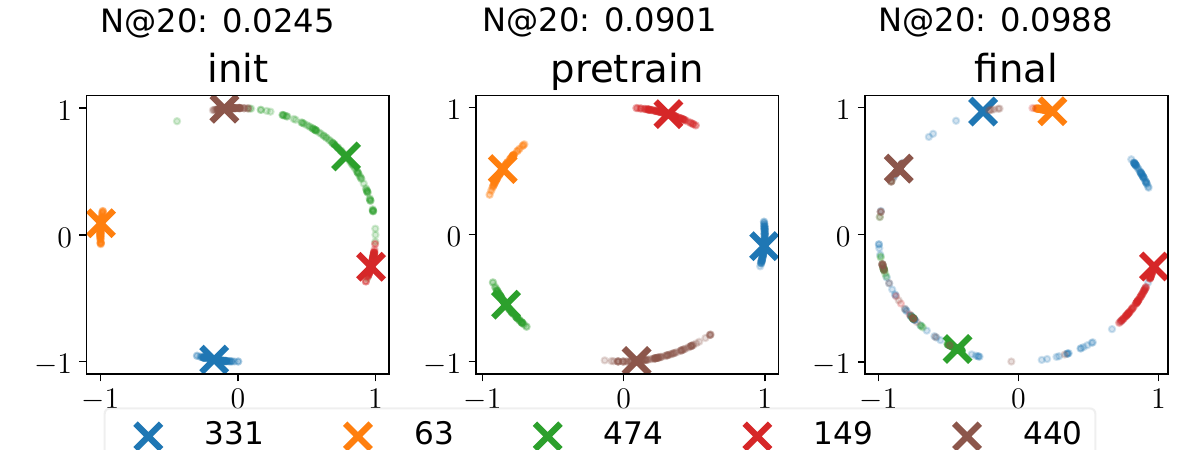}
            \vspace*{-3mm}
            \caption[]%
            { Visualization of sampled entity embeddings and their anchor meta-embeddings. The N@20 value annotated at the top of each plot is the NDCG@20 metric evaluated under the nominated setting. The sampled entities are represented as dots and each cross is a meta-embedding. The same color is used to scatter sampled entities assigned the identical anchor embedding on initialization. Note that we normalize the embeddings in $[-1, 1]$ scale for visualization consistency. }   
            \label{fig:qualitative_gowalla}
    % \vspace*{-4mm}
\end{figure}

\textbf{Case Study on Learned Entity Embeddings.}
Due to the page limit, in this part, we only discuss the case study on Gowalla dataset. To study the relationship between entity embeddings and their anchor embeddings, we sample 5 meta-embeddings and for each meta-embedding, we randomly select 100 entities whose initial anchor embedding is the chosen meta-embedding on initialization. Then we devise T-SNE \cite{van2008visualizing} dimensionality reduction technique to visualize all selected meta-embeddings and entity embeddings in stages of framework initialization, completion of pretrain and final model after assignment weight update. The scatter plots can be found in Figure \ref{fig:qualitative_gowalla}. From the three subplots, we can observe that on initialization, most entity embeddings are scattered around their anchor embeddings closely. 
At this stage, the entity embeddings do not contain any semantic correlations, thus the poor model accuracy. When the model pretrain is completed, meta-embeddings are learned according to the METIS assignment allocation. 
At this stage, entity embeddings spread more evenly around their anchor embeddings due to the freeze of assignment matrix. The recommendation performance has been largely improved owing to the learning of meta-embeddings. Finally, with the assignment weight update conducted, more flexibility is given in generating entity embeddings. 
The entity representations are more diversified and the sampled meta-embeddings are not necessarily the centroid of the entity group labeled the same color. This is because the assignment update process provides the ability to reselect their anchor embeddings, further strengthening the quality of entity embeddings.

\begin{figure}
            \centering 
            \includegraphics[width=.45\textwidth]{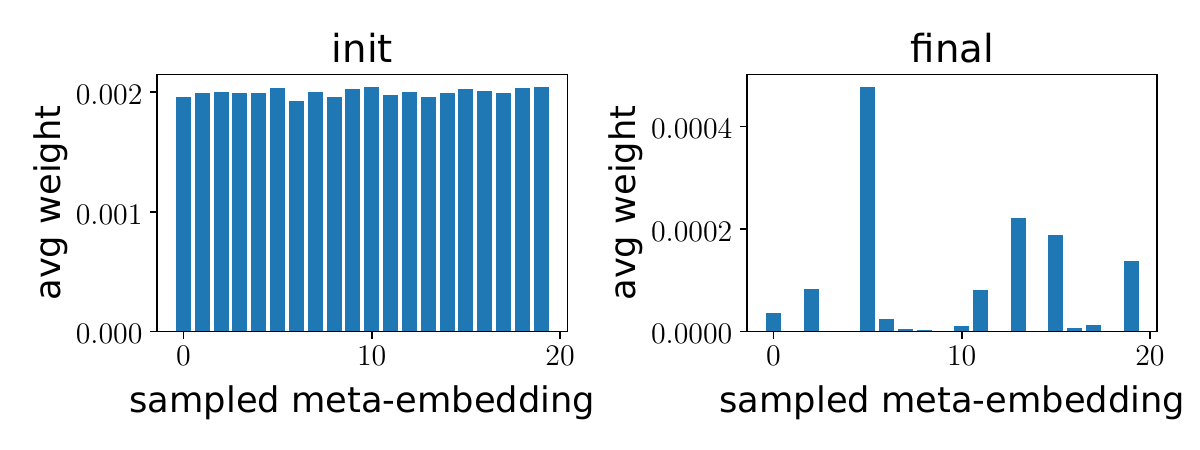}
            \vspace*{-4mm}
            \caption[]%
            { Average assignment weights of sampled meta-embeddings. The weights in the graphs have a small magnitude because we are taking the average weights over all users/items in the dataset.}    
            \label{fig:assignment_gowalla}
        % \vspace*{-4mm}
    
\end{figure}

\textbf{Case Study on Learned Assignments.}
To study the change of assignment weight distribution before and after the weight update process, we sampled 20 meta-embeddings and plotted the average assignment weight distribution graphs for the initial and final stages of the model training in Figure \ref{fig:assignment_gowalla}. A huge weight shift can be spotted after multiple pseudo-inverse-based update rounds were conducted on the METIS initialized assignment matrix. Initially, METIS assigns roughly the same number of entities to each meta-embedding. This can be verified as the average weights for all sampled meta-embedding are similar. In the final stage, only a few outliers can be found in the average weight charts. In contrast, all other meta-embeddings have extremely small average weight values, meaning during the weight update process, critical meta-embeddings can be learned gradually, and more entities are assigned to these critical ones as the anchor embedding. 

\subsection{Hyperparameter Analysis (RQ3)}
\begin{figure*}[t!] 
    \centering
    \begin{subfigure}[b]{.35\textwidth}
        \includegraphics[width=\textwidth]{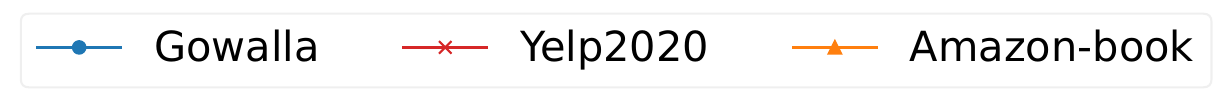}
        %  \caption[]%
        % {{\small test2}}
    \end{subfigure}%
    \vfill
    \begin{subfigure}[b]{.245\textwidth}
        % \centering
        \includegraphics[width=.9\textwidth]{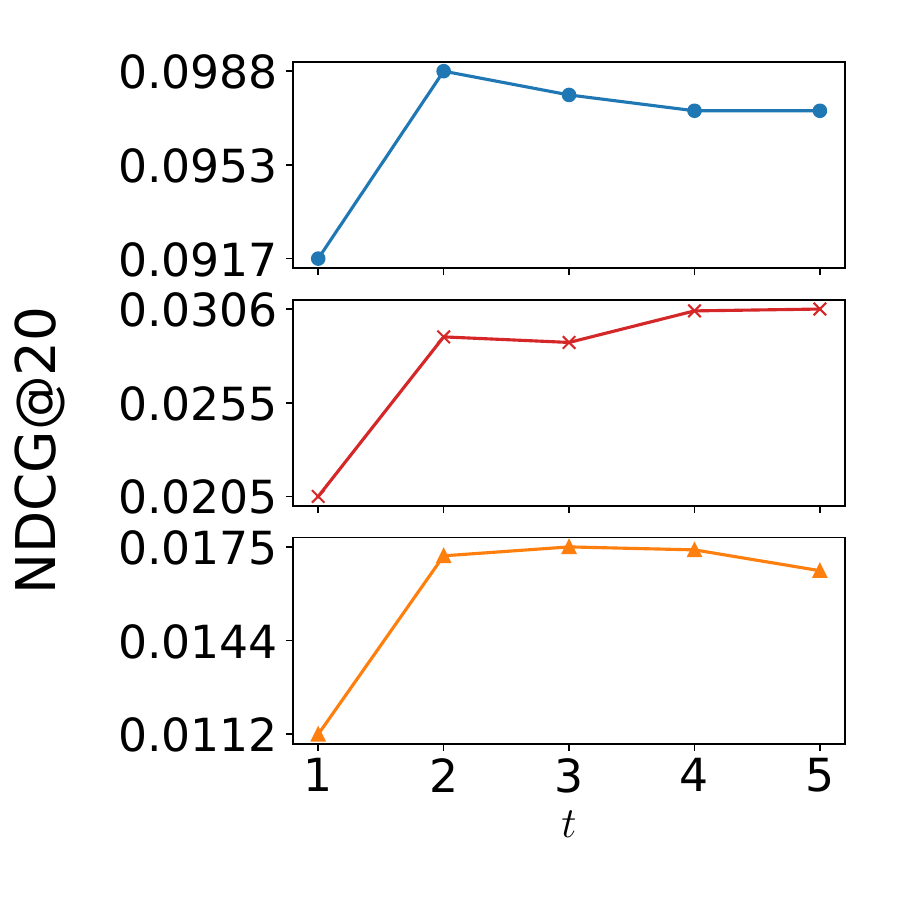}
        \vspace*{-3mm}
        \caption[]%
        {{\small Effect of $t$}}
        \label{fig:comp_embs_per_entity}
    \end{subfigure}%
     \begin{subfigure}[b]{.245\textwidth}
        % \centering
        \includegraphics[width=.9\textwidth]{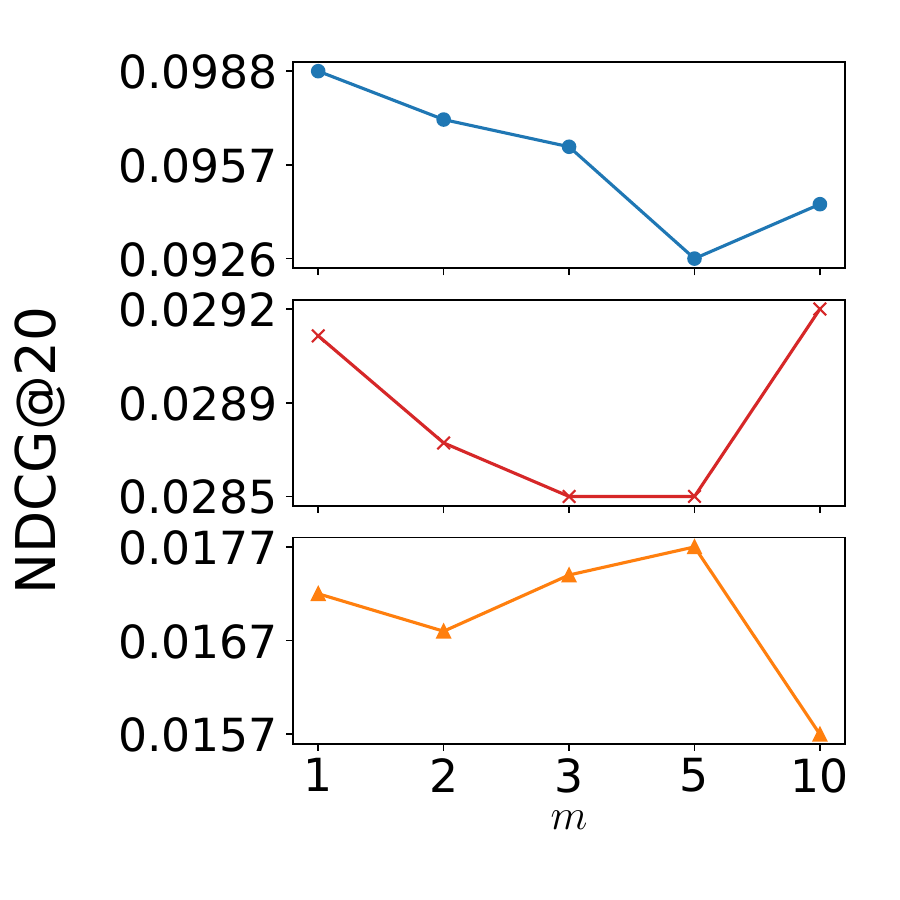}
        \vspace*{-3mm}
        \caption[]%
        {{\small Effect of $m$}}
        \label{fig:update_frequency}
    \end{subfigure}%
    \begin{subfigure}[b]{.245\textwidth}
        % \centering
        \includegraphics[width=.9\textwidth]{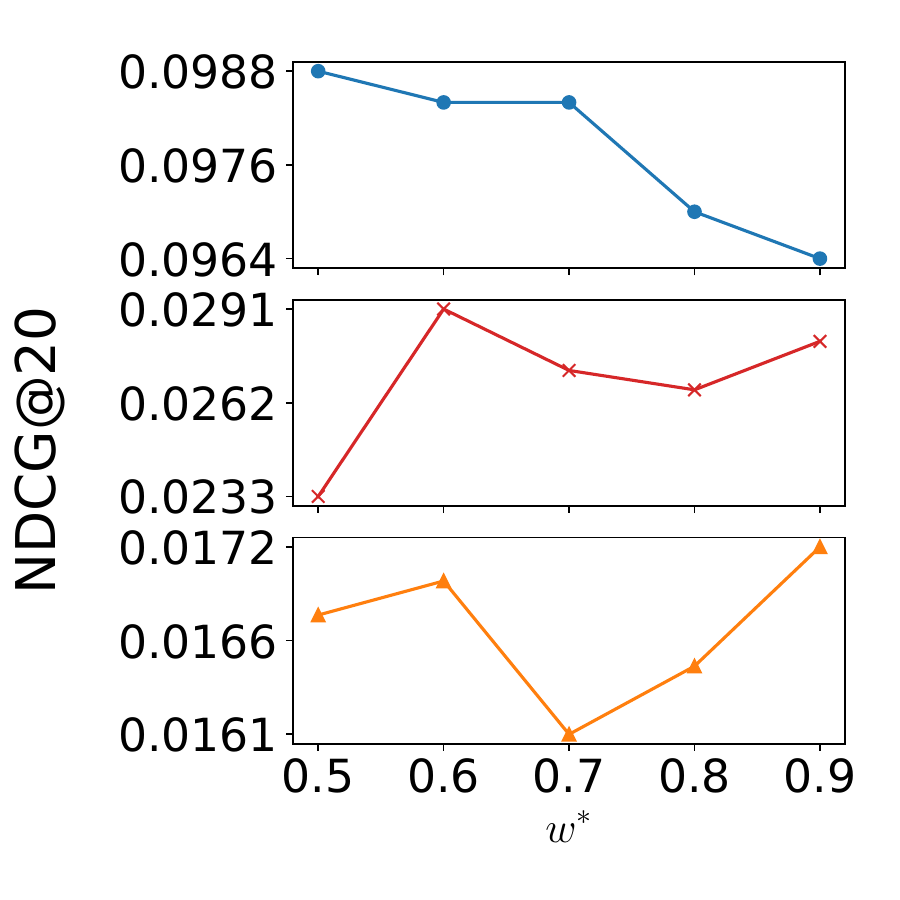}
        \vspace*{-3mm}
        \caption[]%
        {{\small Effect of $w^*$}}
        \label{fig:init_anchor_weight}
    \end{subfigure}%
    \begin{subfigure}[b]{.245\textwidth}
        % \centering
        \includegraphics[width=.9\textwidth]{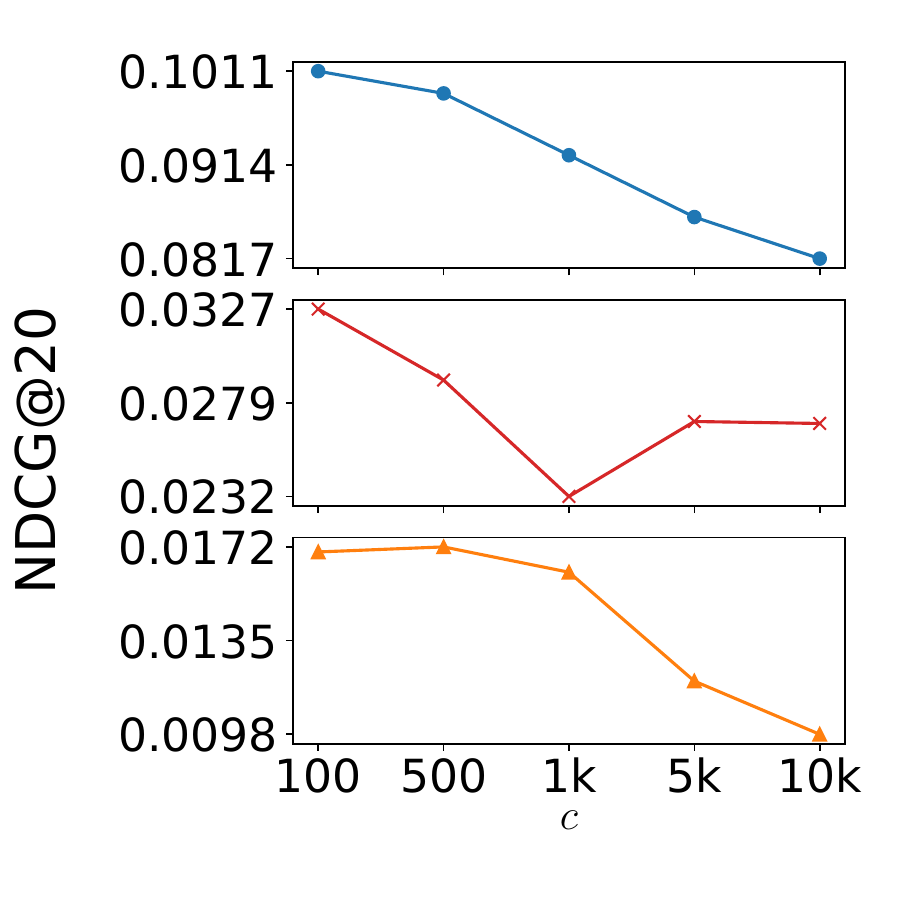}
        \vspace*{-3mm}
        \caption[]%
        {{\small Effect of $c$}}
        \label{fig:bucket_size}
    \end{subfigure}%
    \vspace*{-4mm}
    \caption{The performance of LEGCF on w.r.t. various hyperparameter settings.}
    \label{fig:hyperparams}
    \vspace*{-2mm}
\end{figure*}

To study the sensitiveness of our work on hyperparameters, we conduct hyperparameter analysis on the number of meta-embeddings assigned to each entity $t$, the assignment matrix update frequency $m$, the initial weight of the anchor meta-embedding $w^*$ as well as the bucket size of the meta-embedding codebook $c$.

\textbf{Number of Meta-embeddings $t$ per Entity.}
The recommendation performance regarding the number of compositional embeddings assigned to each entity $t$ is plotted in Figure \ref{fig:comp_embs_per_entity}. The set of $t's$ for testing is $\{1, 2, 3, 4, 5\}$. It is spotted that when each entity is assigned only 1 meta-embedding, the recommendation performance is severely impacted on all three datasets. This is because when each entity embedding is only generated using one meta-embedding, the meta-embedding mapping scheme is indifferent from single hash mapping, which causes serious hash collision.
This statement can be verified as the change of $t$ from 1 to 2 on both datasets leads to a great leap in performance immediately. 

\textbf{Assignment Matrix Update Frequency} $q$. We alter the assignment matrix update frequency $q$ to update every epoch and every 2, 3, 5 and 10 epochs respectively. The recommendation performance regarding different update frequencies is shown in Figure \ref{fig:update_frequency}. It is witnessed that updating the assignment matrix once every epoch in general provides a (nearly) optimal performance result for all three datasets. For Yelp2020, LEGCF is insensitive to the assignment weight update frequency $q$ as the change of update frequency only causes a maximum performance difference of 2.4\%. For Amazon-book dataset, changing the value of $q$ may cause a performance difference of 11.3\%. Our experiments indicate that the best assignment update frequency varies case by case. Some datasets require frequent updates to learn the weights of composed meta-embeddings in time, while some other datasets desire multiple meta-embedding learning epochs between two assignment update processes to learn expressive meta-embeddings. 

\textbf{Initial Weight of the Anchor Meta-embedding} $w^*$. The chart showing the relationship between model performance and the initial weight of anchor meta-embedding is depicted in Figure \ref{fig:init_anchor_weight}. The set of $w^*$ values for testing is $\{0.5; 0.6; 0.7; 0.8; 0.9\}$. It is observed that for Gowalla dataset, the best result is computed using the initial anchor weight of $0.5$. As the value of $w^*$ increases, a 2.4\% drop in performance is caused. The performance of Yelp2020 and Amazon-book w.r.t. the value of $w^*$ follows the same trend, that $w^* = 0.6$ yield satisfactory results, but further increasing $w^*$ will first cause the decrease in performance, and then the performance is gradually improved again when $w^*$ rises to $0.8$ and $0.9$.

\textbf{Meta-embedding Bucket Size} $c$. We draw the meta-codebook buckets $c$ from the set $\{100; 500; 1,000; 5,000; 10,000\}$ for analysis. The graphs in this section are shown in Figure \ref{fig:bucket_size}.
It is witnessed that for all three datasets, LEGCF attains the best performance under the small bucket size $c$ settings. For Gowalla and Yelp2020, the best bucket size is $100$ and for Amazon-book, the best bucket size is $500$. As the bucket size increases to more than $1,000$, its performance becomes less competitive. This phenomenon is especially obvious on the largest dataset Amazon-book, as the increase in bucket size leads to a rapid drop in performance. Our results indicate that LEGCF is designed to work under tight memory budget.

\subsection{Framework Generalizability (RQ4)}
\begin{table}[t!]
\renewcommand{\arraystretch}{.88}

% \vspace*{-4mm}
\caption{Performance comparison between various GNN base recommenders. The best performance under each setting is indicated with bold font.}
\vspace*{-4mm}
\centering
\setlength\tabcolsep{1.3pt}

\label{tab:generalizability}
\begin{tabular}{l|cc|cc|cc}
%\hline
\toprule
\multicolumn{1}{c|}{}       & \multicolumn{2}{c|}{Gowalla}                             & \multicolumn{2}{c|}{Yelp2020}     & \multicolumn{2}{c}{Amazon-book}                       \\ \hline
\multicolumn{1}{c|}{Setting} & N@20                       & R@20                        & N@20                       & R@20     & N@20                       & R@20                   \\ \hline
LightGCN - LEGCF            & \textbf{0.0988}            & \textbf{0.1444}             & \textbf{0.0291}            & \textbf{0.0548}     &  \textbf{0.0172} & \textbf{0.0259}      \\
LightGCN - UD               & \multicolumn{1}{c}{0.0584} & \multicolumn{1}{c|}{0.0863} & \multicolumn{1}{c}{0.0169} & \multicolumn{1}{c|}{0.0321} & \multicolumn{1}{c}{0.0074} & \multicolumn{1}{c}{0.0110} \\ \hline
NGCF - LEGCF                & \textbf{0.0797}            & \textbf{0.1092}             & \textbf{0.0328}            & \textbf{0.0602}   & \textbf{0.0130} & \textbf{0.0204}         \\
NGCF - UD                   & \multicolumn{1}{c}{0.0547} & \multicolumn{1}{c|}{0.0810} & \multicolumn{1}{c}{0.0115} & \multicolumn{1}{c|}{0.0230} & \multicolumn{1}{c}{0.0062} & \multicolumn{1}{c}{0.0096} \\ \hline
LR-GCCF - LEGCF             & \textbf{0.0716}            & \textbf{0.1052}             & \textbf{0.0157}            & \textbf{0.0272}   & \textbf{0.0126} & \textbf{0.0192}         \\
LR-GCCF - UD                & \multicolumn{1}{c}{0.0001} & \multicolumn{1}{c|}{0.0002} & \multicolumn{1}{c}{0.0002} & \multicolumn{1}{c|}{0.0005} & \multicolumn{1}{c}{0.0006} & \multicolumn{1}{c}{0.0010} \\ 
%\hline
\bottomrule
\end{tabular}

\end{table}
We examine the generalizability of our framework in adopting various GNN-based recommenders. To do this, apart from LightGCN, we select two other popular GNN base recommenders, namely, the NGCF \cite{wang2019neural} and the LR-GCCF \cite{chen2020revisiting} to conduct performance testing. We apply appropriate  dimension aggregation techniques 
to feed their propagated embeddings to our assignment update process. We keep all other settings identical to LightGCN for consistency. In addition, we also implement the same GNN recommender settings on unified dimension (\textbf{UD}) embedding tables of similar parameter sizes for comparison. We report the performance in Table \ref{tab:generalizability}.

From the table, it is noted that LEGCF works well on all three GNN recommenders. LightGCN is the strongest base recommender for Gowalla and Amazon-book datasets. For Yelp2020 dataset, NGCF performs slightly better than LightGCN. 
Contrasting LEGCF's strong ability in devising GNN to reinforce embedding quality under the memory-restrained scenario, \textbf{UD} settings are severely affected by the choice of the GNN recommenders on all three datasets. This is verified when comparing the performance between LEGCF and unified dimension settings under a tight memory budget. It is discovered that with the help of our method, the three base recommenders boost the performance substantially. Meanwhile, for unified dimension settings, the recommendation performance is heavily affected by the number of usable parameters. One extreme case is when the LR-GCCF is utilized to train unified dimension embedding tables for the three datasets, the performance is completely obsolete.

\section{Conclusion} \label{sec:conclusion}
In this paper, we propose a novel graph-based compositional embedding framework LEGCF to resolve the challenges of fixed hash mapping and unanimous weight allocation to all composed meta-embeddings by introducing a compact meta-embedding codebook and a learnable assignment matrix. In LEGCF, we compute GNN propagated compositonal embeddings for entities. We propose an efficient assignment update strategy to reflect the meta-embedding usage for composition in time. Both innovative designs significantly improve the quality of composed entity embeddings and boost the recommendation performance. Our comprehensive experiment indicates LEGCF's superiority in performance over other embedding optimization work under memory-restrained scenarios.

\section*{Acknowledgments}
This work is partially supported by the Australian Research Council under the streams of Future Fellowship (Grant No. FT210100624), Discovery Early Career Researcher Award (Grant No. DE230101033), and Discovery Project (Grants No. DP240101108, and No. DP240101814).

\newpage
\bibliographystyle{ACM-Reference-Format}
\bibliography{customized}

%%
%% If your work has an appendix, this is the place to put it.
% \appendix
% \input{sections/appendices.tex}

\end{document}